\documentclass[11pt,twoside]{article}

\usepackage{graphicx}
\usepackage{asp2006}
\usepackage{epsf}
\usepackage{psfig}
\usepackage{lscape}

\begin{document}
\title{The Current Understanding on the UV Upturn}
\author{Sukyoung K. Yi}
\affil{Yonsei University, Department of Astronomy, Seoul 120-749, Korea}

\begin{abstract}
The unexpected high bump in the UV part of the spectrum found in nearby giant
elliptical galaxies, a.k.a. the UV upturn, has been a subject of debate.
A remarkable progress has been made lately from the observational side, mainly
involving space telescopes. The GALEX UV telescope has been obtaining
thousands of giant ellipticals in the nearby universe, while HST is resolving
local galaxies into stars and star clusters. An important clue has
also been found regarding the origin of hot HB stars, and perhaps of
sdB stars. That is, extreme amounts of helium are suspected to be the origin of the extended HB and even to the UV upturn phenomenon.
A flurry of studies are pursuing the physics behind it.
All this makes me optimistic that the origin of the UV upturn will
be revealed in the next few years.
I review some of the most notable progress and remaining issues.
\end{abstract}

\section{Introduction}

A review on the UV upturn phenomenon may usually start with a following or similar definition:
``a bump in the UV spectrum between the Lyman limit and 2500\AA\, is found
{\em virtually in all bright spheroidal galaxies}'' (e.g., Yi \& Yoon 2004).
This seems no longer true!
While earlier studies based on a small sample of nearby galaxies
led us to think so, a much greater sample from the recent GALEX database
appears to disprove it. Only a small fraction of elliptical galaxies
show a strong UV upturn and it is generally limited to the brightest
cluster galaxies (Yi et al. 2005).
This review is about the recent development on this seemingly-old topic.
I recycle some of the contents in my earlier review given in the first Hot
Subdwarf and Related Objects workshop held in Keele, UK (Yi \& Yoon 2004).
For a more traditional review, readers are referred to the articles of
Greggio \& Renzini (1999) and O'Connell (1999).

\section{Previous observations}

The UV upturn has been a mystery ever since it was first found by
the OAO-2 space telescope (Code \& Welch 1979). According to the
opacity effect more metal-rich populations show redder colours,
and hence giant elliptical galaxies were not expected to
contain any substantial number of hot stars to show a UV upturn.
Yet, it was confirmed by subsequent space missions, ANS (de Boer 1982),
IUE (Bertola et al. 1982) and HUT (Brown et al. 1997).
Figure 1 shows an example spectrum of the giant elliptical galaxy NGC\,4552
mosaicked from multi-band measurements.

Some of the observational findings based on the nearby bright
elliptical galaxies are particularly noteworthy.
The positive correlation between the UV-to-optical colour
(i.e., the strength of the UV upturn) and the Mg2 line strength
found by Burstein et al. (1987) through IUE observations has urged
theorists to construct novel scenarios in which old ($\ga$ a few Gyr) metal-rich ($\ga Z_{\odot}$)  populations become UV bright
(Greggio \& Renzini 1990; Horch et al. 1992; Dorman et al. 1995).
Also interesting was to find using HUT that,
regardless of the UV strength, the UV spectral slopes at
1000--2000\AA\, in the six UV bright galaxies were similar
suggesting a very small range of temperatures of the UV
sources in these galaxies (Brown et al. 1997), which corresponds
to $T_{\rm eff} \approx 20,000 \pm 3,000$\,K.
In fact, the characteristic temperature of the UV sources seems
strangely somewhat $lower$ in a $stronger$ UV-upturn galaxy (Yi et al. 1998).

\begin{figure}
\begin{center}
\includegraphics[scale=0.4,angle=-0]{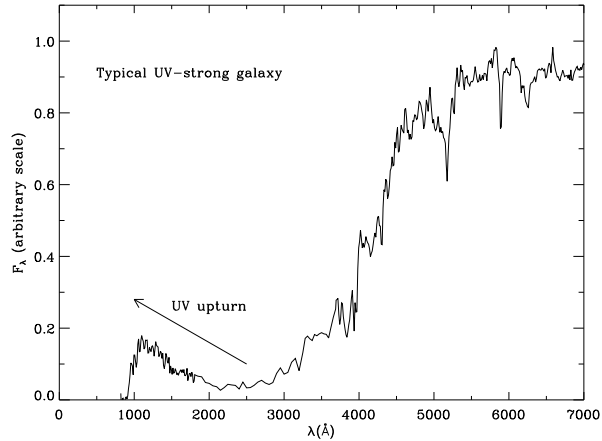}
\end{center}
\caption{The composite spectrum of the giant elliptical galaxy NGC 4552
shows a classic example of the UV upturn. The mosaic spectrum is
originated from HUT (FUV), IUE (NUV), and ground-based telescope (optical).
Excerpted from Yi, Demarque, \& Oemler (1998).
}
\label{fig1}
\end{figure}

\section{Theory}

Theorists aim to present a model that explains three basic observational facts:
\begin{enumerate}
\item UV upturn being present in bright elliptical galaxies
\item the positive correlation between the strength of the UV upturn and the
$optical$ metal line (Mg2) strength, and
\item a narrow range of temperature of UV sources.
\end{enumerate}

Young stars are difficult to satisfy these facts and thus thought unlikely
to be the main driver of the UV upturn.
The focus has been on how an old population can develop hot stars.
Post-AGB stars (central stars of planetary nebulae) are too short-lived
and more fatally too hot most of their lifetime, hence violating $item$ 3.
There is a good consensus that hot (low-mass) horizontal-branch (HB)
stars are the more natural candidates. Here I introduce two classical
solutions based on the HB hypothesis.

\subsection{Metal-poor HB hypothesis}

It is widely known that metal-poor HB stars can be hot and make
good UV sources when they are old (e.g., Lee et al. 1994).
Thus, the first scenario was naturally that an order of 20\% of
the stellar mass of bright elliptical galaxies are extremely old and metal-poor
populations (Park \& Lee 1997). The strength of this scenario is
that the oldest stars in a galaxy are likely the most metal-poor
and to be in the core, where the UV upturn is found to be strong.
In this scenario, the UV vs Mg2 relation does not present any causality
connection but simply a result of tracing different populations in
terms of metallicity. Mg2 is exhibited by the majority metal-rich stars
while the UV flux is dominated by the old metal-poor stars.
The narrow range of temperature is easily explained as well.
On the other hand, the mass fraction of order $\sim 20$\% is too high
by the standard galactic chemical evolution theory.
Canonical models suggest the metal-poor fraction of $\la 10$\%.
If metal-poor stars are present at such a high level,
there must also be a large number of intermediate-metallicity
(20--50\% solar) stars, which will make galaxy's integrated
metallicity too low and integrated colours too blue, compared
to the observed values.
Moreover, the age of the oldest stars, i.e.
the main UV sources, is required in this scenario to be 20--30\%
older than the average Milky Way globular clusters (Yi et al. 1999).
This would pose a big challenge but there may be a rescue (see \S 4).

\subsection{Metal-rich HB hypothesis}

Through a gedanken experiment Greggio \& Renzini (1990) noted a
possibility that extremely low-mass HB stars may completely skip the
AGB phase and dubbed it ``AGB Manq\'{u}e stage''. Through this stage
metal-rich populations could become UV bright.
This is particularly effective for a high value of helium abundance
(Dorman et al. 1995). If galactic helium is enriched
with respect to heavy elements at a rate of $\Delta Y$/$\Delta Z
\ga 2.5$ this means that the stage would be very effective in galaxy
scales as well (Horch et al. 1992). It could be similarly effective
if the mass loss rate in metal-rich stars is 30--40\% higher than that of
metal-poor stars (Yi et al. 1997a). Either of the two conditions
would be sufficient while they can also complement each other.
Both of these conditions are difficult to validate empirically
but plausible (Yi et al. 1998).
In this scenario, metal-rich stars may become UV bright in two
steps: (1) they lose more mass on the red giant phase due to the
opacity effect and become low-mass HB stars, and (2) extremely
low-mass HB stars stay in the hot phase for a long time and
directly become white dwarfs, effectively skipping the red,
asymptotic giant phase (Yi et al. 1997a, 1997b). This scenario
reproduces most of the features of the UV upturn (Bressan et al.
1994; Yi et al. 1998). The UV vs Mg2 relation is naturally explained
as a UV vs metallicity relation.
However, its validity heavily hinges upon the purely-theoretical
(and hence vulnerable to criticisms) late-stage stellar evolution
models of metal-rich stars.

\subsection{Metal-poor or metal-rich HB?}

Both of these scenarios are equally appealing but their
implications on the age of bright elliptical galaxies are
substantially different.
The metal-poor hypothesis suggest UV-upturn galaxies are 30\%
older than Milky Way and requires the universe to be
older than currently believed, suggesting a large cosmological
constant. The metal-rich hypothesis on the other hand suggests
that elliptical galaxies are not necessarily older than the Milky Way halo.

\begin{figure}[!t]
\begin{center}
\includegraphics[scale=0.4,angle=-0]{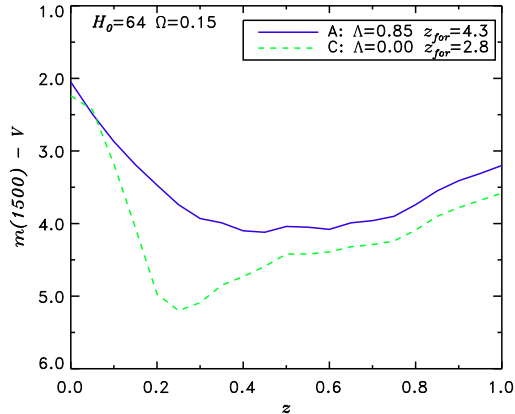}
\end{center}
\caption{
The two classic models (Model A: metal-poor HB, Model C: metal-rich HB)
predict different evolution history. While precise calibrations are difficult,
the UV developing pace is in general predicted to be faster for more metal-rich
populations. Excerpted from Yi et al. (1999).
}
\label{fig2}
\end{figure}

\section{Issues}

Readers may get an impression by reading the previous sections
that we have solid and successful theories. Quite contrarily,
there are several critical issues to be understood before we can
ever claim so.

\subsection{$\alpha$-enhancement}

Theorists (including myself) often interpret the UV vs Mg2
relation as a metallicity effect on the UV flux. However, it
should be noted that Mg2 strength may not be representative of the
overall metallicity. In fact, it has been known that elliptical
galaxies are enhanced in $\alpha$-elements with respect to iron.
We then naturally wonder if it is not the overall metallicity but
$\alpha$-enhancement that generates the UV upturn. To perform this
test, we need $\alpha$-enhanced stellar models. The $Y^2$
Isochrones group have released their $\alpha$-enhanced 
stellar models for the main sequence (MS) through red giant branch (RGB) 
(Kim et al. 2002). But, no $\alpha$-enhanced HB
models are publicly available yet. $\alpha$-enhancement can have several
impacts on the galaxy spectral evolution. First, it changes the
stellar evolutionary time scale, as CNO abundance affects the
nuclear generation rates. Second, it changes opacities and thus
the surface temperatures of stars. These two effects will make a
change in the mass loss computed using a parameterised formula,
such as the Reimers (1975) formula. For a fixed mass loss efficiency, we
find the $\alpha$-enhanced ([$\alpha$/Fe]=0.3--0.6) tracks yield
$\approx 0.03 M_{\odot}$ smaller mass loss at ages 5--8Gyr but
$\approx 0.03 M_{\odot}$ greater mass loss at ages
$\ga 8$\,Gyr, compared to the standard ([$\alpha$/Fe]=0)
tracks. $\alpha$-enhancement must have similar opacity effects on
the HB evolution, while its effect on the mass loss on the HB should be negligible.
Thus its effects are expected to be greater to the MS to RGB than
to the HB phase. Considering this, I have decided to inspect the
overall effects of $\alpha$-enhancement by just adopting new
$\alpha$-enhanced MS through RGB tracks, ignoring the change in
the HB models. My earlier review (Yi \& Yoon 2004) shows the
results for two metallicities and three values of $\alpha$-enhancement.
It can be summarised as follows.
In old metal-poor models $\alpha$-enhancement causes a positive effect to the
$relative$ UV strength because (1) it causes a slight increase in
mass loss on the RGB and (2) it causes MS stars and red giants to
be redder and fainter in $V$ band. The [$\alpha$/Fe]=0.3 model
roughly reproduces the SED of a typical UV-strong metal-poor
globular cluster, which is satisfying. The metal-rich models
on the other hand do not show any appreciable change in response to
$\alpha$-enhancement. Because giant elliptical galaxies are largely
metal-rich (roughly solar) and the light contribution from
metal-poor stars is not substantial, it is unlikely for
$\alpha$-enhancement to play a major role to the UV upturn.

\subsection{EHB stars in star clusters}

With the HST spatial resolution, a number of studies have found
hot, extended horizontal branch (EHB) stars in globular clusters
(e.g., Piotto et al. 1999). They are efficient UV sources and
important candidates for the main UV sources in elliptical
galaxies; but canonical population synthesis models have difficulty
reproducing them as they are observed (number density,
colours and brightness).

NGC 6791 is a particularly interesting case.
This old (8-9Gyr) metal-rich (twice solar) open cluster is unique 
resembling the stellar populations of the giant elliptical galaxies.
Strikingly, 9 out of its 32 seemingly-HB stars have the properties
of typical EHB stars (Kaluzny \& Udalski 1992; Liebert et al.
1994), while canonical models do not predict any (Yong et al. 2000).
It is critical to understand the origin of these
{\em old hot metal-rich stars}.
Landsman et al. (1998), based on UIT data, concluded that NGC 6791, if observed
from afar without fore/background stellar contamination, would
exhibit a UV upturn just like the ones seen in elliptical galaxies.

Through detailed synthetic HB modelling we found that it is
impossible to generate an HB with such a severely-bimodal colour
distribution as shown in this cluster, unless an extremely (and
unrealistically) large mass dispersion is adopted. In the hope of
finding a mechanism that produces such an HB Yong et al. (2000)
explored the effect of mass loss $on$ the HB. Yong et al. found
that with some mass loss taking place on the HB ($\approx
10^{-9}-10^{-10}\,M_{\odot}\,yr^{-1}$) HB stars born cool quickly
become hot, suggesting that mass loss on the HB might be an
effective mechanism of producing such stars.
Vink \& Cassisi (2002) however pointed out that the level of the
mass loss assumed by Yong et al. is too high to justify in their
radiation pressure calculations in the context of single-star evolution. Green et al. (2000)
reported that most of these hot stars in NGC 6791 are in binary
systems. If they are close binaries and experience mass transfer
it would be an effective mechanism for mass loss. 
But at the moment it is difficult to conclude whether
binarity had causality on their EHB nature or not.

\subsection{Binaries}

SdB/O stars, the central objects of this conference, may be the
field counterparts of the EHB stars in clusters.
They have the properties similar to those of the UV sources in the
UV-upturn galaxies. Surprisingly, more than 70\% of sdB
stars are found to be in binary systems (Saffer et al. 2000;
Maxted et al. 2001).

Han et al. (2003) used a binary population synthesis technique to
study the effects of binary evolution and found that 75--90\% of
sdB stars should be in binaries. SdBs are detected to be in a
small mass range centred at 0.5\,$M_{\odot}$, but Han et al.
found that the range should be in truth as wide as 0.3 through 0.8
$M_{\odot}$. They predict a birthrate of 0.05~$yr^{-1}$ for
Population I stars and 6 million sdB stars in the disc. Assuming
the Galactic Disc mass of 5x$10^{10}$\,$M_{\odot}$, this means
roughly 100 sdB stars per $10^6 M_{\odot}$. In a
back-of-the-envelope calculation, there are roughly a few thousand
HB stars per million solar mass in globular cluster populations.
A comparison between the sdB rate (100 per $10^6 M_{\odot}$) and
that of the HB (say, 5000 per $10^6 M_{\odot}$) suggests that an old
disc population may develop 1 sdB star for 50 HB stars (2\%). This
sounds by and large reasonable from the EHB-to-HB number ratio
found in globular clusters.
But it is hardly impressive from the perspective of searching
for copious UV sources in galaxies.
For comparison, NGC 6791 has roughly 30\% (8 EHB-like stars
out of 32 HB-like stars) and the UV-brightest Galactic globular cluster
$\omega$\,Cen has 20\%. These two examples show an order of magnitude
higher values of EHB-to-HB ratio than deduced from a simple estimation
based on the binary population synthesis models.
Yet, even $\omega$\,Cen does not exhibit a UV upturn as observed
in giant elliptical galaxies: $FUV$$-$$V$ is comparable but $FUV$$-$$NUV$ 
is 1--2 magnitudes redder than found in ellipticals.
If this calculation is
realistic at least within an order, binary mass transfer may not be
sufficient to provide the origin of the majority of the UV sources in
UV-upturn galaxies. On the other hand, a larger sdB production rate might be
plausible in elliptical galaxy environment due to large age and/or
large metallicity.

A considerably more detailed investigation was presented by Han et al. (2007).
They constructed the population synthesis models including binaries
of varied properties (in mass ratio and separation).
The conclusions from their prediction can be summarised as
(1) most of the UV light of ellipticals comes from binary sdB stars
(2) a UV upturn starts to appear as early as when the galaxy is 1.5Gyr old
(3) and the $FUV$$-$$V$ colour stays virtually constant since then.
This is an important prediction because this is the first study that
realistically consider binary products in population models.
One immediately notices that the item (3) contradicts the single-star
population models of Yi et al. (1999) discussed in \S 3.3 and Figure 2.

\subsection{Other issues}

There are other important issues as well.  For example, the
late-stage flash mixing scenarios and the like (D'Cruz et al.
1996; Brown et al. 2001) may also be effective ways of producing
hot stars (such as sdB stars) in old populations. Their typical
temperature range ($T_{\rm eff} \gg 20,000$\,K) 
and the predicted birthrate may not be entirely
consistent with the UV upturn shown in elliptical galaxies, however.

Another important observational constraint comes from the HST UV
images of M32. First, Brown et al. (2000) found that PAGB stars
are two orders of magnitudes fewer than predicted by simple
stellar evolution theory. This is significant as PAGB stars are
thought to account for 10--30\% of the UV flux in the UV-upturn galaxies
(Ferguson \& Davidsen 1993). More
importantly, they find too many $faint$ hot HB stars to reproduce with
standard population models that are based on the mass loss rate
calibrated to the globular cluster HB morphology.
It is possible to reproduce the observed number of hot
stars in M32-type populations if a greater mass loss rate is used,
which would be consistent with the variable mass loss hypothesis
(Willson et al. 1996; Yi et al. 1997b, 1998).
But theoretical justification is a problem again.

\section{GALEX observations}

The single star population synthesis models (\S 3.3) predict
a rapid decline in $FUV$$-$$V$ with increasing redshift (lookback time),
while the binary models suggest no significant change.
This stark contrast provides an important test.

GALEX is NASA's UV space telescope mission that can do just this.
It has sensitive FUV and NUV detectors and reaches passive (no star formation)
old populations (such as many elliptical galaxies) out to $z \sim 0.2$
(Martin et al. 2005).
Its Deep Imaging Survey (DIS) is obtaining the UV images of tens of
galaxy clusters using $\ga 20,000$ seconds of exposure.
The UV upturn is found to be the strongest in the
brightest cluster elliptical galaxies (BCGs) and hence we have tried to obtain
accurate photometric data on the BCGs in our galaxy cluster sample.
Besides, a number of lower-redshift ($z \sim 0.1$) BCGs have been sampled
from the shallower Medium-deep Imaging Survey (MIS) as described in
Schawinski et al. (2007).
The UV photometry turned out to be very tricky because there are many
background UV sources that are not easily identifiable in shallow images.
The background confusion would easily cause underestimation on
the UV brightness. Occasionally, small foreground objects that are
invisible in the optical images contaminate the UV flux of our target
galaxy as well.

From the up-to-date GALEX database, Ree et al. (2007) obtained the data
for seven BCGs from DIS and five from MIS.
A small fraction of the BCGs had star formation signatures (Yi et al. 2005)
and hence had to be removed from our sample.
Figure 3 shows the look-back time evolution of the apparent (not $K-$corrected) $FUV - V$ colour for the BCGs at $z <$ 0.2.
The FUV flux fades rapidly with redshift. The colours are derived from total
magnitudes to minimize aperture effect. Model lines are calibrated to
the colour range ($FUV - V = 5.4 - 6.4$) of the giant elliptical galaxies
in nearby clusters ($open~circles$), and passively evolved and redshifted
with look-back time so that they can be directly compared with the
observed data of the BCGs ($filled~circles$)
in GALEX DIS (black) and MIS (grey) mode. The size of circle symbols
represents the absolute total luminosity in $r$-band. The solid and
dashed lines are from the passively evolving UV-to-optical spectra
of the ``metal-poor'' and ``metal-rich'' HB models (\S 3).
The regions filled with oblique lines denote the predicted colour
range from these two extreme models. The dotted line indicates
the apparent colour expected when the local UV upturn galaxy NGC~1399
model spectrum is redshifted without the effect of stellar evolution.
The binary population models would be similar to the non-evolving model.
The higher redshift data points at 0.33 and 0.6 are the HST data
from Brown et al. (2000, 2003)
The model fits by Ree et al. (2007) and Lee et al. (2005a) suggest
that {\em the GALEX data show a UV flux decline with
lookback time at the rate $\Delta (FUV-V)/\Delta t = 0.54$ mag/Gyr.}
Although a definite answer requires more data, the current sample
seems more consistent with the prediction from the single-star population
models.
Any population model aiming to explain the UV upturn phenomenon would be
obliged to reproduce this unique data.

\begin{figure}[t]
\begin{center}
\includegraphics[scale=0.6,angle=0]{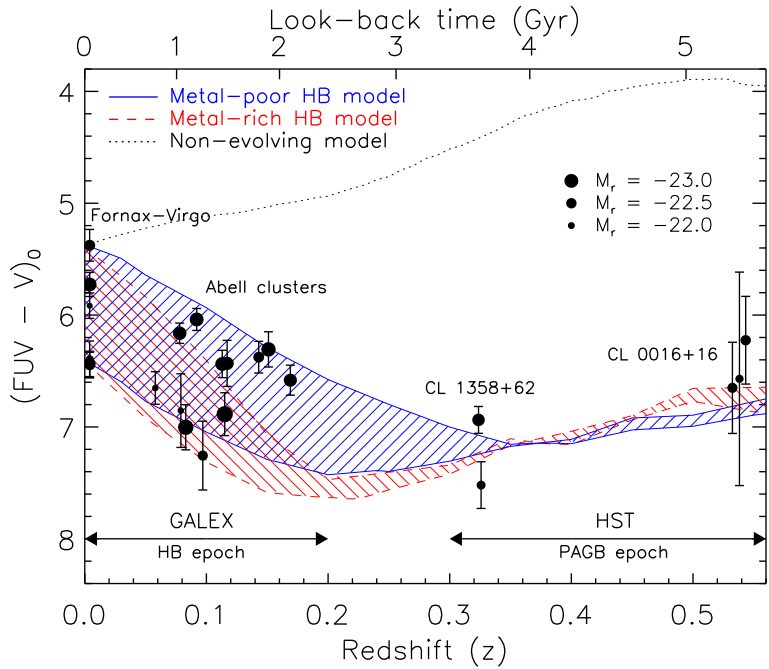}
\end{center}
\caption{
Look-back time evolution of the apparent (not $K-$corrected) $FUV - V$
colour for the brightest cluster elliptical galaxies (BCGs) at $z \la$ 0.2.
FUV flux fades rapidly with redshift which is consistent with the
prediction from the single-star population models (\S 3.3).
See text for details. Excerpted from Ree et al. (2007).
}
\label{fig3}
\end{figure}

\section{New issue: enhanced-helium population}

\begin{figure}[t]
\begin{center}
\includegraphics[scale=0.6,angle=-90]{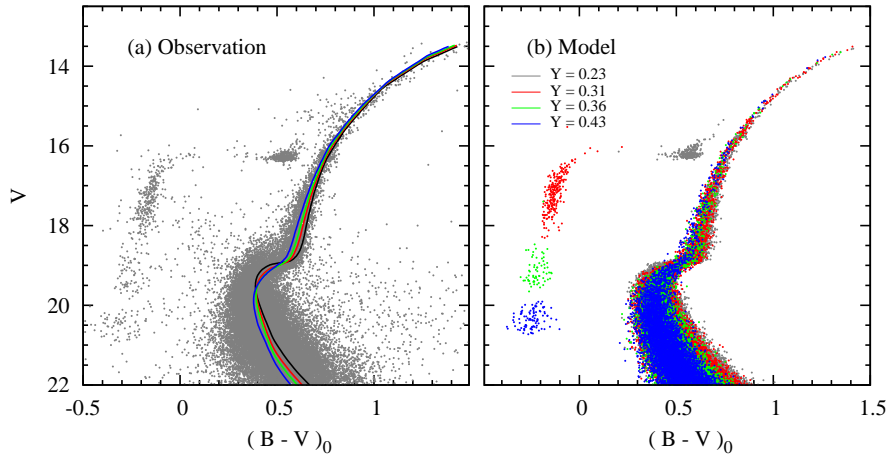}
\end{center}
\caption{The observed and modeled colour-magnitude diagrams of
the globular cluster NGC\,2808. $left$: The cluster shows an exceptionally
wide distribution of horizontal branch stars. $right$: It can be
precisely reproduced by theory for example by assuming a large range of helium
abundance. Excerpted from Lee et al. (2005b)
}
\label{fig4}
\end{figure}

A remarkable new information has recently emerged.
Observations for the colour-magnitude diagrams on globular clusters
$\omega$ Cen and NGC\,2808 revealed the multiple nature of their stellar
populations. The most massive globular cluster $\omega$ Cen for example
is now known to have up to four different metallicities both for the
main sequence and the red giant branch (Anderson 2002; Bedin et al. 2004).
Most shockingly, the bluest main sequence is found spectroscopically 
to be more metal-rich (Ferraro et al. 2004) which implies an extremely
high helium abundance of $Y \approx 0.4$.
Interestingly, Lee et al. (2005b)
noted that such a helium-rich stellar population
would evolve into extremely hot HB explaining the hitherto mysterious
origin for the EHB stars of $\omega$ Cen. 
Lee et al. claims that the same phenomenon is seen in NGC\,2808 as well.
Such a high helium abundance could in fact be more mysterious than
the origin of the EHB stars itself, hence became a hot topic.
The high value of helium abundance ($Y \approx 0.4$) seems particularly
impossible when it is combined with its low metallicity empirically constrained
($Z \approx 0.002$--0.003).
This leads to $\Delta$$Y$/$\Delta$$Z \approx 70$ which is extremely
unlikely from the galactic chemical enrichment point of view
unless some exotic situation is at work, such as the chemical inhomogeneity
in the proto-galactic cloud enriched by first stars (Choi \& Yi 2007).

No matter what the physical process may be, the CMD fits unanimously
suggest that the high value of helium is the easiest solution.
Figure 4 shows Lee et al. (2005b)'s
comparison between the observed and model CMDs
assuming that the hot end of the HB morphology is primarily governed
by the variation in the helium abundance. The reproduction is impressive.
If the EHB is indeed produced by helium variation, then, it almost seems
that we are going back to two decades ago in terms of the debate on the
second parameter for the HB morphology (see Lee et al. 1994).
According to Lee et al. (2007), a pronounced EHB is more easily found among
more massive globular clusters, which forces us to think deeply about the nature of globular clusters in general.

It is not yet clear whether the enhanced helium interpretation is
physically plausible and whether it can be similarly significant to the galaxy
scale where for example the primordial chemical fluctuation proposed
should be hidden in the mean properties of the stellar populations of a
galaxy (see Choi \& Yi 2007).

\section {New issue: UV-bright globular clusters in M87}

\begin{figure}[t]
\begin{center}
\includegraphics[scale=0.6,angle=0]{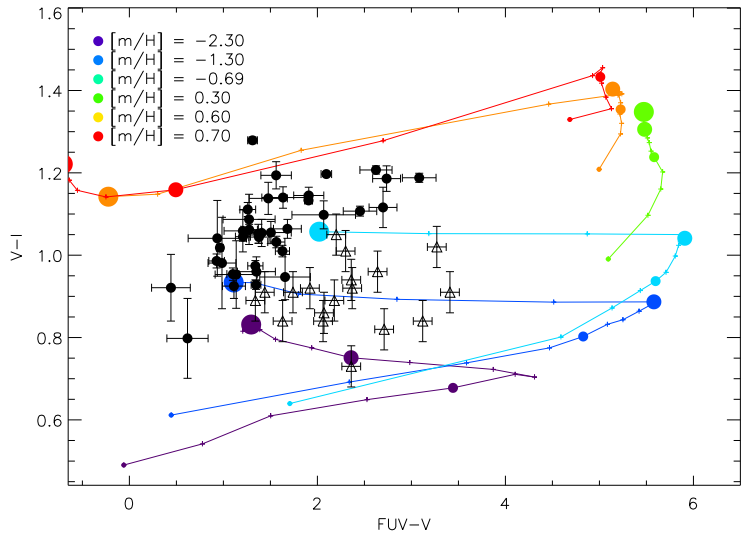}
\end{center}
\caption{
Model (FUV − V) versus (V−I) grid for a range of metallicities and ages,
generated from stellar models with the fiducial value of He enrichment
(ΔY/ΔZ = 2). The lowest age plotted is 1 Gyr and the largest age
plotted is 15 Gyr. Ages 1, 5, 10 and 15 Gyr are shown using filled circles
of increasing sizes. The globular cluster data of M87 (filled circles)
and Milky Way (open triangles) with errors are overplotted.
It is apparent that the M87 photometry lies outside the age range
1–-14 Gyr for all metallicities. Excerpted from Kaviraj et al. (2007).}
\label{fig5}
\end{figure}

The discovery of numerous UV-bright globular clusters in the giant
elliptical galaxy M87 is also remarkable (Sohn et al. 2006).
Using HST/STIS UV filters Sohn et al. found 66
globular clusters from small fields of view
most of which are bluer and hotter than the Milky
Way counterparts. Kaviraj et al. (2007) found that the canonical
population synthesis models with normal values of helium cannot
reproduce their UV properties at all, as shown in Figure 5.
Kaviraj et al. found that their UV brightness can be reproduced
if a similar amount of EHB stars found in the $\omega$ Cen study
by Lee et al. (2005b) are artificially 
added to the canonical population models as well.
This is very interesting. 
The more massive M87 is believed to contain 2 orders of magnitude
more star clusters than the Milky Way does and thus it is very natural 
for us to find many more UV-bright globular clusters from M87 than from Milky
Way.
This can be compatible with the enhanced helium hypothesis.
If the enhanced helium is present, say in 10\% of the star clusters,
roughly 10 clusters in the Milky Way and up to 1000 clusters in M87
might be helium-enhanced and thus UV-bright. A part of them may have been found by Sohn et al. (2006).

\section{Conclusions}

The UV satellite GALEX is obtaining a valuable UV spectral evolution
data for numerous bright cluster galaxies. The apparent trend in redshift
vs $FUV-V$ colour seems consistent with the prediction from the single
stellar population models.
This is comforting while observers feel obliged to build up their
database much more substantially in order to make it statistically robust.

Two new issues are notable.
Firstly, binary population synthesis community feels odd to find that the
simplistic single-star population models are found to be good enough.
The in-principle more advanced binary population models are obliged to reproduce the observed CMDs of simple populations (globular clusters) before
attempting to model galaxies. For example, I am very eager to see 
their models reproduce the ordinary HB first, before explaining the EHB.

Secondly, the
enhanced helium hypothesis based on the globular clusters in Milky
Way and M87 is a very exciting possibility.
The deduced value of the helium abundance seems unphysical to be
a global property for the galaxy but may be possible for small systems
that are vulnerable to a chemical fluctuation in the proto-galaxy cloud.
While a more detailed investigation is called for it may be difficult
to be influential to the entire stellar population of a galaxy.
For instance, adding all spectral energy distributions of the Milky
Way globular clusters would not yield anything close to the spectrum
of a UV upturn galaxy. Of course, a metallicity difference may act as
an added complication.

The secret will be revealed through time and hard work, perhaps very soon.

\acknowledgements
I thank Uli Heber the Bamberg meeting organiser for the great workshop.
Special thanks go to Chang H. Ree for providing  slides for my
review presentation at the meeting. I thank Chul Chung for generating
Figure 3 specifically for this article. This article is based on
many insightful discussions with Chang H. Ree, Young-Wook Lee, Mike Rich,
Jean-Michel Deharveng, Suk-Jin Yoon, Tony Sohn, Sugata Kaviraj,
Andres Jordan, Kevin Schawinski, and David Brown.
I acknowledge many helps from the GALEX science operation and data analysis
team. This trip and review was possible with the support from the KOSEF fund.

\end{document}